\documentclass[
    ,final            
  ]
  {aipproc}

\layoutstyle{6x9}

\begin{document}

\title{CSPOB - Continuous Spectrophotometry of Black Holes}


\author{S. K. Chakrabarti}{
address={S.N. Bose National Center for Basic Sciences, JD-Block, Salt Lake, Kolkata, 700098}
,altaddress={Indian Centre of Space Physics, chakraba@bose.res.in (Principal Investigator)}
}
\author{D. Bhoumik}{
address={Indian Centre for Space Physics, 43 Chalantika, Garia Station Rd., Kolkata 700098}
}
\author{D. Debnath}{
address={Indian Centre for Space Physics, 43 Chalantika, Garia Station Rd., Kolkata 700098}
}
\author{R. Sarkar}{
address={Indian Centre for Space Physics, 43 Chalantika, Garia Station Rd., Kolkata 700098}
}
\author{A. Nandi}{
  address={Indian Centre for Space Physics, 43 Chalantika, Garia Station Rd., Kolkata 700098}
,altaddress={On deputation from Indian Space Research Organization}
}
\author{V. Yadav}{
address={Indian Centre for Space Physics, 43 Chalantika, Garia Station Rd., Kolkata 700098}
,altaddress={On deputation from Indian Space Research Organization}
}
\author{A. R. Rao}{
address={Tata Institute of Fundamental Research, Homi Bhabha Road, Colaba, Mumbai 400005}
}
\begin{abstract}

The goal of a small and dedicated satellite called the "Continuous 
Spectro-Photometry of Black Holes" or CSPOB is to provide the essential tool for the theoretical 
understanding of the hydrodynamic and magneto-hydrodynamic flows around black holes. 
In its life time of about three to four years, only a half a dozen black holes will be observed 
continuously with a pair of CSPOBs. Changes in the spectral 
and temporal variability properties of the high-energy emission would be caught 
as they happen. Several important questions are expected to be answered and many 
puzzles would be sorted out with this mission.
\end{abstract}

\maketitle


\section{Introduction}

Black holes are the most enigmatic objects in the Universe. Last forty years, several 
space missions have gathered data from these objects, and yet, a true and complete 
understanding of the physical processes around black holes has remained a mystery. 
While the quality and quantity of data have gone up exponentially, they were motivated by general properties
of compact objects and not by specific properties of black holes.
For example, none of them were geared towards capturing the outcome of solutions around black holes. 
because of these even after watching hundreds of objects (in random 
time intervals), the model builders have to explain each object on a case by case basis. Theoreticians
have lesser choice, as they believe that whatever is observed are to be explained 
from solutions of equations with appropriate boundary conditions. 
Thus continuous observations would allow them to refine the equations, adding more terms
and obtain better solutions. Day by day, it becoming clearer from the observational facts
(e.g., Soria et al. 2001; Smith et al. 2001, 2002, 2007; 
Wu et al. 2002;  Shaposhnikov \& Titarchuk, 2006; Shaposhnikov et al. 2007) that accretion flows
have a substantial amount of sub-Keplerian components as was predicted long ago
(Chakrabarti, 1990; Chakrabarti \& Titarchuk, 1995; Chakrabarti, 1996). 

With a pair of CSPOBs (Chakrabarti et al. 2007), for the first time, we will be able to 
answer questions like: How should the standard model of accretion of matter
be modified? What are the roles played by a non-Keplerian or sub-Keplerian flow and how do
they change with time?  How does the variation of the accretion rates of the Keplerian
and sub-Keplerian flow reflected in the light curves, spectral and timing properties?
Can we distinguish between various models of Compton cloud which range
from a magnetic corona to the puffed up post-shock flow called CENBOL?
What is the origin of the fast oscillations (Quasi-Periodic Oscillations or QPOs)
observed at high energies and, why do they occasionally appear with frequencies 
in certain ratios? How important is the magnetic field in the dynamics of the 
accretion flow and the jet? Is there a relationship between iron the
line emission shape and the QPO frequencies? Do we really know where the iron line 
comes from, the disk or the jet? etc.

\section{Components of CSPOB}

The instruments that we plan to fabricate is simple and is solely dedicated to observe only 
half a dozen black holes (unless, of course, some
more exciting transient black holes divert its attention)
in its lifetime. The payloads are restricted to be about 20 kilograms each. Each CSPOB 
will be a 640 cm$^2$ Si-PIN photo-diode based array of detectors which is 
sensitive in the $1-50$ keV range. Large area Si drift detectors are also being
studied as an alternative. Both would be equally good to  have $< 5 \%$ energy resolution.
The data in 256 energy channels will be stored every 100 sec and thus time
resolved spectra would be obtained and time lags/leads in Keplerian and sub-Keplerian 
components, light curve variations of intriguing objects such as GRS 1915+105 etc.
There would be two All Sky Proportional Counters (ASPCs). It will revolve with the satellite 
and scan the sky for new outbursts. Indian Centre for Space Physics is developing Si-PIN 
detectors and also collaborating with other National and International organizations 
to fly this simple tool in near future. A multi-wavelength large scale mission is 
{\it not} suitable for our purpose. To our knowledge no mission has been proposed so far
to observe a single celestial body other than CSPOB if one excludes bodies within our solar system.

\begin{figure}[h]
\includegraphics[height=7truecm]{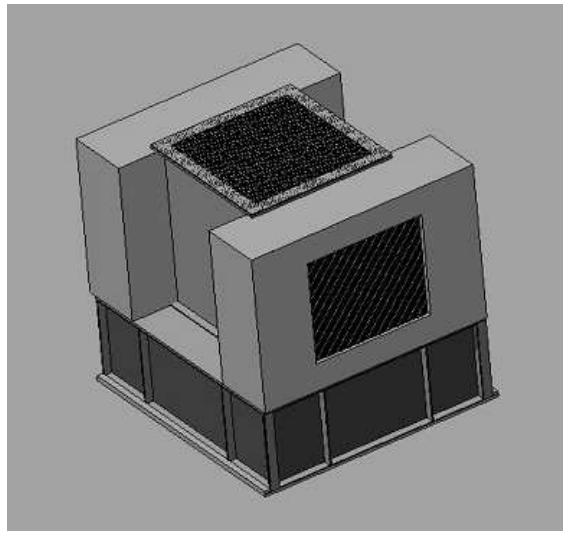}
\caption{A three dimensional view of the complete payload with the Spectrophotometer
at the center and the all sky proportional counters (ASPCs) on both sides.} 
\end{figure}

\begin{figure} [h]
\includegraphics[height=7truecm]{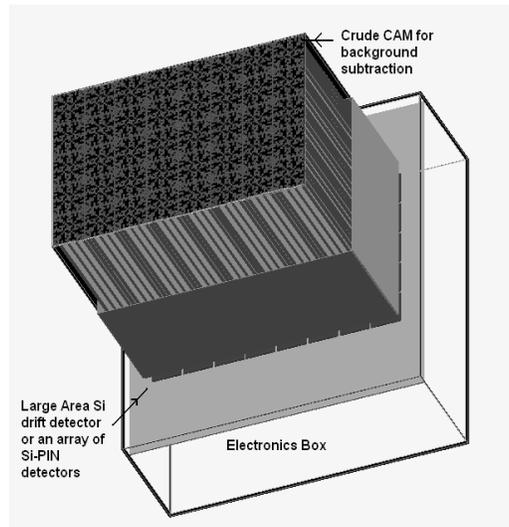}
\caption{The details of the Spectrophotometer with a crude type Coded Aperture Mask (CAM)
for background subtraction, collimator, Si based detector and the electronics box.}
\end{figure}

\begin{figure}[h]
\includegraphics[height=7truecm]{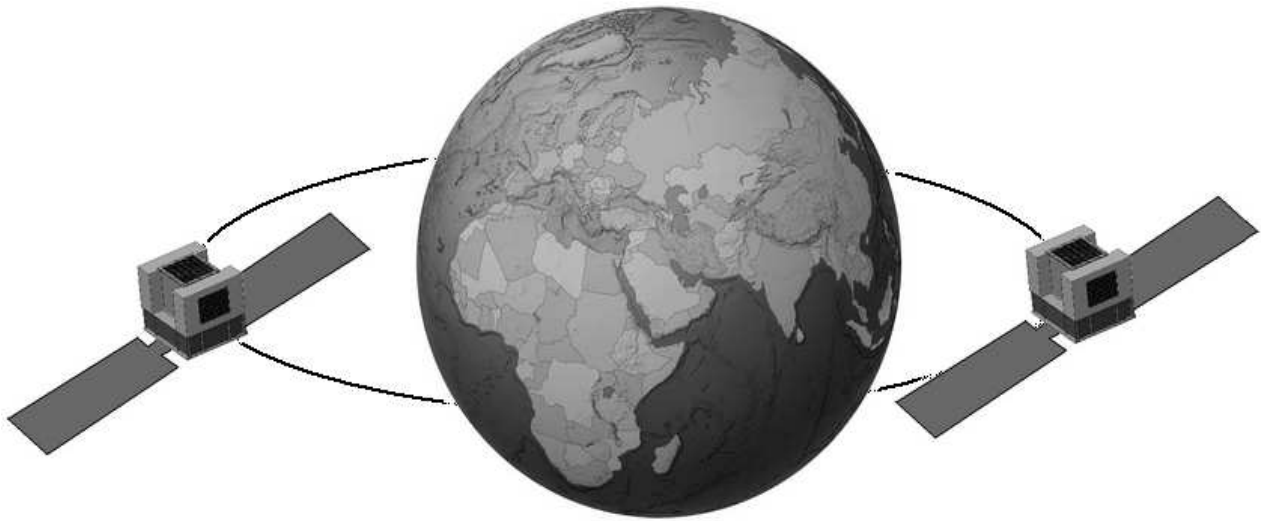}
\caption{A pair of dedicated CSPOBs is essential to keep track of X-ray activities 
close to a compact object}
\end{figure}

\section{Brief Science objectives}

We now briefly summarize the reasons why we believe that continuous observation
is essential. First of all, black holes having Kerr radius $r_k=GM/c^2$ is 
very small, about $15$km for a nano-quasar of with a rapidly spinning black hole of mass $10M_\odot$.
The light crossing time is only $50\mu$s. A black hole accretion being
necessarily transonic (Chakrabarti, 1990) , the flow 
close to the black hole is relativistic. The infall time-scale in a Keplerian flow
vary with viscosity and it can take a few days for the matter to come from the 
outer edge to the inner edge. But the net inflow is known to have a substantial 
amount of almost freely falling sub-Keplerian matter and indeed this determines 
the emitted spectrum and its variability unless the spectrum is 
very very soft. High frequency QPOs suggest that the emitted intensity
does change in a matter of a few tens of milliseconds. The low and intermediate frequency QPOs,
which are more common, change in a matter of seconds or less and are well explained
by rapidly oscillating shocks originating in sub-Keplerian flow. Class transitions in objects
like GRS 1915+105 take place in a matter of few seconds, and their nature change totally
in a matter of a few hours (Chakrabarti et al. 2004; Choudhury et al. this volume). 
Some of these involve luminosity variations by orders of magnitude. Thus
it is obvious that all these can happen only if the local effects are important
which are triggered by sub-Keplerian flows. A constant vigil is required to really understand
how the accretion rates of the Keplerian and sub-Keplerian flows change with time. Constant 
observation of black body and power-law components would give the variation of geometry
close to the inner edge since the power-law component is produced from the interception of the
soft-components. 
  
\section{Conclusions}

We justify the goal to observe continuously celestial objects other 
than the Sun and present a viable mechanism to do a thorough study using a pair of
small satellites. We require to observe both the soft and the hard components 
of the galactic black hole candidates. We are studying the feasibility
of the Si-PIN based photo-diodes (Bhoumik et al. this volume) as well as 
Si drift detectors for this purpose. Details will be presented elsewhere.


\begin{theacknowledgments}
D. Debnath acknowledges the support of CSIR through a NET fellowship.
\end{theacknowledgments}

\end{document}